\documentclass[anonymous,10pt,conference]{IEEEtran}
\IEEEoverridecommandlockouts
\usepackage{cite}
\usepackage{amsmath,amssymb,amsfonts}
\usepackage{algorithmic}
\usepackage{graphicx}
\usepackage{textcomp}
\usepackage[hidelinks]{hyperref}
\usepackage{xcolor}
\usepackage{balance}
\usepackage{listings}
\usepackage{booktabs,tabularx,multirow,threeparttable,makecell}

\newcommand{\M}[1]{{\color{black}#1}}

\newcommand{\DSD}{\textit{Desiview}}
\newcommand{\ECR}{\textit{Desiview4FT}}
\newcommand{\DR}{\textit{DRC}}
\newcommand{\AECR}{\textit{Desiview4FA}}

\def\BibTeX{{\rm B\kern-.05em{\sc i\kern-.025em b}\kern-.08em
    T\kern-.1667em\lower.7ex\hbox{E}\kern-.125emX}}
\begin{document}

\title{Distilling Desired Comments for Enhanced Code Review with Large Language Models
}

\author{\IEEEauthorblockN{Yongda Yu}
\IEEEauthorblockA{\textit{Software Institute} \\
\textit{Nanjing University}\\
Nanjing, China \\
yuyongda@smail.nju.edu.cn}
\and
\IEEEauthorblockN{Lei Zhang}
\IEEEauthorblockA{\textit{Software Institute} \\
\textit{Nanjing University}\\
Nanjing, China \\
522023320200@smail.nju.edu.cn}
\and
\IEEEauthorblockN{Guoping Rong*}
\IEEEauthorblockA{\textit{Software Institute} \\
\textit{Nanjing University}\\
Nanjing, China \\
ronggp@nju.edu.cn}
\and
\IEEEauthorblockN{Haifeng Shen}
\IEEEauthorblockA{\textit{Faculty of Science and Engineering} \\
\textit{Southern Cross University}\\
Bilinga, Queensland, Australia \\
haifeng.shen@scu.edu.au}
\and
\IEEEauthorblockN{Jiahao Zhang}
\IEEEauthorblockA{\textit{Software Institute} \\
\textit{Nanjing University}\\
Nanjing, China \\
211250031@smail.nju.edu.cn}
\and
\IEEEauthorblockN{Haoxiang Yan}
\IEEEauthorblockA{\textit{Software Institute} \\
\textit{Nanjing University}\\
Nanjing, China \\
211250009@smail.nju.edu.cn}
\and
\IEEEauthorblockN{Guohao Shi}
\IEEEauthorblockA{\textit{Software Institute} \\
\textit{Nanjing University}\\
Nanjing, China \\
211250033@smail.nju.edu.cn}
\and
\IEEEauthorblockN{Dong Shao}
\IEEEauthorblockA{\textit{Software Institute} \\
\textit{Nanjing University}\\
Nanjing, China \\
dongshao@nju.edu.cn}
\and
\IEEEauthorblockN{Ruiqi Pan}
\IEEEauthorblockA{\textit{Huawei Technologies Co., Ltd.} \\
Shenzhen, China \\
panruiqi@huawei.com}
\and
\IEEEauthorblockN{Yuan Li}
\IEEEauthorblockA{\textit{Huawei Technologies Co., Ltd.} \\
Shenzhen, China \\
liyuan50@huawei.com}
\and
\IEEEauthorblockN{Qiushi Wang}
\IEEEauthorblockA{\textit{Huawei Technologies Co., Ltd.} \\
Shenzhen, China \\
wangqiushi6@huawei.com}
\and
\IEEEauthorblockN{Zhao Tian}
\IEEEauthorblockA{\textit{Huawei Technologies Co., Ltd.} \\
Shenzhen, China \\
tianzhao@huawei.com}
}

\maketitle

\begin{abstract}
There has been a growing interest in using Large Language Models (LLMs) for code review thanks to their proven proficiency in code comprehension. The primary objective of most review scenarios is to generate desired review comments (\DR{s}) that explicitly identify issues to trigger code fixes. However, existing LLM-based solutions are not so effective in generating \DR{s} for various reasons such as hallucination. To enhance their code review ability, they need to be fine-tuned with a customized dataset that is ideally full of \DR{s}. Nevertheless, such a dataset is not yet available, while manual annotation of \DR{s} is too laborious to be practical. In this paper, we propose a dataset distillation method, \DSD, which can automatically construct a distilled dataset by identifying \DR{s} from a code review dataset. Experiments on the CodeReviewer dataset comprising more than 150K review entries show that \DSD\ achieves an impressive performance of 88.93\%, 80.37\%, 86.67\%, and 84.44\% in terms of Precision, Recall, Accuracy, and F1, respectively, surpassing state-of-the-art methods. To validate the effect of such a distilled dataset on enhancing LLMs' code review ability, we first fine-tune the latest LLaMA series (i.e., LLaMA 3 and LLaMA 3.1) to build model \ECR. We then enhance the model training effect through KTO alignment by feeding those review comments identified as non-\DR{s} to the LLMs, resulting in model \AECR. Verification results indicate that \AECR\ slightly outperforms \ECR, while both models have significantly improved against the base models in terms of generating \DR{s}. Human evaluation confirms that both models identify issues more accurately and tend to generate review comments that better describe the issues contained in the code than the base LLMs do.

\end{abstract}

\begin{IEEEkeywords}
LLM, Automated Code Review, Fine-tuning, Alignment
\end{IEEEkeywords}

\section{Introduction}\label{sec:intro}
Code review is a crucial component of modern software development and has been widely applied in the development of software systems~\cite{gousios2014exploratory}. The primary objective of most review scenarios is to generate review comments that explicitly identify issues in a code to trigger code fixes before it is executed for quality assurance~\cite{lu2023llama, kononenko2016code}. We refer to these comments as \DR{s} (Desired Review Comments). Typically, a \DR\ should accurately pinpoint the locations of the issues in the code, correctly describe the nature of the issues, and/or lead to meaningful subsequent repairs to the code. However, as code review is generally a lengthy and costly process~\cite{kononenko2016code,bosu2013impact}, considerable efforts have been made to automate the process by adopting machine learning or deep learning techniques~\cite{li2022codereviewer}. In recent years, the emergence of Large Language Models (LLMs) has introduced new possibilities to automated code review~\cite{lu2023llama}. Owing to their stronger semantic understanding capabilities than traditional machine learning methods and general language models, they have the potential to enable more accurate identification of subtle issues in the code. Additionally, their inherent content generation capabilities allow them to generate better review comments~\cite{li2022codereviewer, lu2023llama}.

\begin{figure*}[ht]
    \centering
    \includegraphics[width=7.2in]{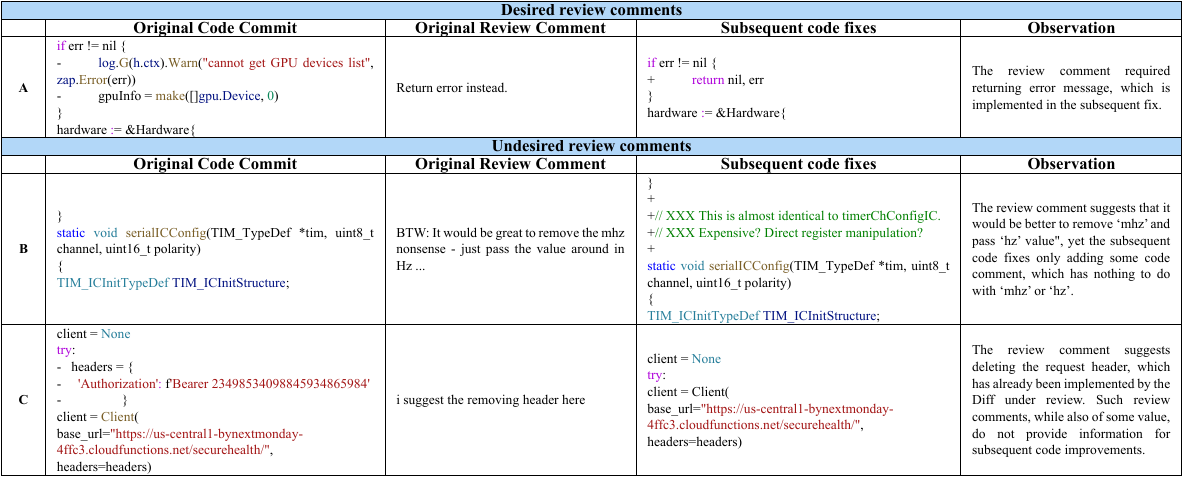}
    \caption{Examples of desired and undesired review comments in CodeReviewer\cite{li2022codereviewer} dataset}
    \label{fig:BadExample}
\end{figure*}

\M{
However, existing LLM-based solutions may not be able to effectively generate \DR{s}
for various reasons such as
their inherent characteristic of hallucination~\cite{hou2023large}. Among these reasons, a critical one is that they are not effectively fine-tuned for the code review task~\cite{fan2023large} and a common cause is the lack of an adequate fine-tuning dataset comprising of \DR{s}~\cite{lu2023llama}. For example, the dataset may contain a considerable proportion of non-\DR\ data (i.e., undesired review comments, cf. Table \ref{tab:DatasetResult}).  }
Figure~\ref{fig:BadExample} illustrates examples of both desired and undesired review comments, which were drawn from one of the commonly used datasets in code review research~\cite{li2022codereviewer}. Example A represents a \DR\ as it identifies an issue in the Diff to be reviewed, which has been fixed in the subsequent code. \M{Conversely, Examples B and C depict undesired review comments, as the subsequent code changes indicate that they do not seem to be triggered by these comments.}
It has been commonly acknowledged that the adequacy of datasets impacts the training effect of LLMs~\cite{zhou2024lima, liu2024coachlm, rejeleene2024towards}. As such, to enhance an LLM's code review ability, it needs to be fine-tuned with a customized dataset that is ideally full of \DR{s}. 

An intuitive way to obtain such a dataset is through manual annotation~\cite{ouyang2022training, mcaleese2024llm}. However, the enormous labor cost (e.g., the dataset~\cite{li2022codereviewer} contains more than 150000 review entries) behind manual annotation~\cite{wang2022self, zhao2023survey} makes it rather impractical. On top of that is its varying quality, which has been repeatedly raised in multiple studies~\cite{wang2022self,luo2024semi,plank2022problem}. Code review researchers thereby have made various attempts to construct such a customized dataset automatically, most of which have relied on simple keyword or rule-based filtering methods~\cite{bosu2015characteristics, li2022codereviewer}. For instance, some studies employ the 10-line rule, which deems a review desired if it results in modifications within 10 lines in the new version of the code~\cite{bosu2015characteristics}. Other studies consider the first record of all review comments as desired by excluding the original author's comments~\cite{li2022codereviewer}. As these methods lack semantic understanding and analysis of both the review comments and the relevant code, their effectiveness is suboptimal such that there is no guarantee that the customized dataset contains a high proportion of \DR{s}. 

This paper aims to fill this gap by proposing a dataset distillation method, \DSD, which can automatically construct a distilled dataset that fine-tunes LLMs for code review tasks by identifying \DR{s} from a code review dataset. By employing this method to distinguish between desired and undesired review comments and subsequently constructing a distilled dataset with a high proportion of \DR{s}, we first fine-tune the latest LLaMA series (i.e., LLaMA 3 and LLaMA 3.1) to build a model \ECR\ and then KTO-align the model to build an enhanced model \AECR. The main contributions of this paper are summarized below.
\begin{itemize}
\item We propose the \DSD\ method for automatically distilling \DR{s} from a code review dataset. It achieves an accuracy of 86.67\% on the CodeReviewer dataset~\cite{li2022codereviewer}, surpassing previous methods including the GPT-4o's 76.50\%.
\item We develop two code review models \ECR\ and \AECR\ by fine-tuning and KTO-aligning the latest LLaMA series with the distilled dataset. Both models have significantly improved against the base models in terms of generating \DR{s} on the CodeReviewer dataset.
\item We conduct a human evaluation of the generated review comments. The results indicate that both \ECR\ and \AECR\ identify issues more accurately and tend to generate review comments that better describe the issues contained in the code than the base LLMs do.
\end{itemize}

\M{The rest of the paper is organized as follows.
Section \ref{sec:related} introduces some related work. Section \ref{sec:method} describes the research methodology followed by the evaluation process in Section \ref{sec:eval}. Section \ref{sec:discussion} discusses the implications, followed by the validity risks in Section \ref{sec:threat}. Section \ref{sec:conclusion} concludes the paper with a summary of contributions and future work.
}

\section{Related work}\label{sec:related}
In this section, we describe related work to our study, including automated code review and applications of LLMs in software engineering.
\subsection{Automated code review}

Code review, as an essential process in software development, has garnered widespread attention from researchers \cite{bosu2013impact, sadowski2018modern}. Given that code review may consume a significant amount of reviewers' effort and time \cite{kononenko2016code, sadowski2018modern}, researchers have increasingly focused on building automated review systems to assist reviewers. An automated review system typically comprises two components: defect detection and review comment recommendation/generation.

Defect detection is used to find potential issues contained in the code snippets under review. For example, DACE \cite{shi2019automatic} uses CNN and LSTM techniques to extract Diff features from the code, thereby predicting the quality of code Diff patches. Some pre-trained models also have been used to assess code quality, such as CodeBert \cite{feng2020codebert} and CodeT5 \cite{wang2021codet5}. CodeBert \cite{feng2020codebert} is a bimodal pre-training model designed for programming languages and natural language. It performs well in tasks such as natural language-based code search and code documentation generation. CodeT5 \cite{wang2021codet5} leverages a unified framework to support both code understanding and generation tasks, thereby facilitating multi-task learning. This method exhibits superior performance compared to previous techniques in several relevant tasks such as code understanding \cite{feng2020codebert} and generation \cite{radford2019language}.

Review comment recommendation/generation produces review comments through retrieval or generation methods. For example, CommentFinder \cite{hong2022commentfinder} uses deep learning techniques to retrieve relevant code review comments, thereby reducing the time reviewers spend writing review comments. DCR \cite{gupta2018intelligent} learns the similarity between code commit Diffs and review comments to retrieve review comments related to a specific code commit. CodeReviewer \cite{li2022codereviewer} achieves notable results in code defect detection, code review comment generation, and code repair tasks by constructing pre-training tasks targeted at code review in an end-to-end manner. LLaMA-Reviewer \cite{lu2023llama} introduces LLMs into code review tasks, using low-parameter fine-tuning techniques to fine-tune LLaMA, achieving impressive results in review comment generation. It is worth noting that the above two studies use the same dataset \cite{li2022codereviewer} for model training and verification. They assume the existence of review comments indicates ground truth without considering whether the review comments actually pertain to the code fixes.

\subsection{Large language models for software engineering}
Recent years have witnessed widespread applications of LLMs in various software engineering tasks, especially in those related to code. For example, CodeLLaMA \cite{roziere2023code}, an LLM by fine-tuning LLaMA2 \cite{touvron2023llama} with a large amount of source code, 
achieves good performance on various code tasks. DeepSeek Coder \cite{guo2024deepseek} is pre-trained on 2 trillion tokens across more than 80 programming languages, surpassing CodeLLaMA in code tasks. StarCoder 2 \cite{lozhkov2024starcoder}, trained on 3.3 to 4.3 trillion tokens with carefully selected data, outperforms the 33B parameter DeepSeek Coder using 15.5B parameters. LLaMA3 \cite{llama3modelcard}, one of the latest versions of the most widely used LLM architecture in the open-source community, has achieved state-of-the-art in multiple tasks. In general, there are three main technical routes for applying LLMs in software engineering -- prompt engineering, fine-tuning, and alignment. 

\textbf{Prompt engineering} focuses on leveraging the inherent capabilities of large models by carefully constructing prompts and implementing processes to achieve better performance. For example, CodeT \cite{chen2022codet} uses prompt engineering to first guide the large model to generate test code corresponding to the code generation task and then continuously verifies the accuracy of the generated code using the test code, thereby achieving higher code generation accuracy. MapCoder \cite{islam2024mapcoder} employs prompt engineering to construct multi-agent prompting, simulating the cycle of recalling relevant examples, planning, code generation, and debugging in the human development process, achieving state-of-the-art in multiple evaluation sets. 

\textbf{Fine-tuning} involves training LLMs with data so that they can solve problems based on the given information without providing examples. Magicoder \cite{wei2024magicoder} enhances the instruction code generation capability of LLMs through fine-tuning by constructing diverse instruction data for code generation, surpassing ChatGPT in code generation performance on the humaneval dataset using a 7B model. LLaMA-Reviewer\cite{lu2023llama} enhances the review capability of LLMs by fine-tuning them with the CodeReviewer \cite{li2022codereviewer} dataset, achieving state-of-the-art in code review tasks. RepairLLaMA \cite{silva2023repairllama} fine-tunes the LLaMA series models to endow them with automatic repair capabilities, achieving state-of-the-art on two code repair datasets. Research has shown that the quality of fine-tuning datasets significantly affects the performance of LLMs \cite{zhou2024lima}. Obtaining higher quality datasets has become an important research direction in fine-tuning LLMs \cite{wei2024magicoder,luo2023wizardcoder}.

\textbf{Alignment} enhances the ability of LLMs to generate valid answers while reducing the probability of generating invalid answers by training them with both desired and undesired datasets \cite{ji2023ai}. Large model alignment algorithms are mainly divided into online alignment and offline alignment. Online alignment algorithms involve online sampling, online scoring, and using the scores to optimize the model. Offline alignment algorithms, on the other hand, optimize model performance using only given desired and undesired data. Online alignment algorithms typically consume a lot of resources but usually perform better, while offline alignment algorithms are the opposite \cite{tang2024understanding}. RLHF \cite{ouyang2022training} is the most representative online alignment method, successfully learning human preferences through a reward model and teaching these preferences to LLMs using the PPO algorithm \cite{schulman2017proximal}. Due to the high resource consumption of online alignment algorithms, researchers have turned their attention to offline alignment algorithms. DPO \cite{rafailov2024direct} is the first proposed offline alignment algorithm, using the LLM itself as the reward model, achieving low-cost alignment of LLMs. However, DPO requires paired data, meaning that a single effective alignment data entry must contain both desired and undesired data under one instruction, which is difficult to obtain in practice. To solve this problem, researchers proposed the KTO \cite{ethayarajh2024kto} alignment method, which does not require paired data for alignment and can also perform alignment in situations where the ratio of desired to undesired data is unbalanced. 

Many researchers have started using alignment algorithms to improve the performance of software engineering tasks. For example, RLSQM (Reinforcement Learning from Static Quality) \cite{steenhoek2023reinforcement} proposes a novel technique to construct a quality model based on human analysis and optimize the LLM using PPO, surpassing GPT-4 in test code generation tasks. StepCoder \cite{dou2024stepcoder} scores code based on feedback from the compiler and uses alignment algorithms to enhance the code generation capability of LLMs, achieving state-of-the-art results on test data. PanGu-Coder2 \cite{shen2023pangu} proposes a Rank Responses to align Test \& Teacher Feedback framework based on alignment technology, effectively improving the performance of LLMs in code generation tasks. Similarly, alignment technology usually requires high-quality data by distinguishing between desired and undesired data to improve model performance. This data is often manually annotated \cite{steenhoek2023reinforcement} or generated based on relevant standards \cite{dou2024stepcoder, shen2023pangu}. Construction of high-quality alignment data remains one of the important research directions in the research of alignment techniques \cite{shen2023large}, and, to the best of our knowledge, there is no such work for code review tasks.

\section{Methodology}\label{sec:method}
The primary objective of this research is to develop an LLM-based solution that is effective in generating \DR{s} for code review tasks. The pivotal component of our research methodology is to construct a customized dataset that \M{contains a high proportion of \DR{s}} with a novel dataset distillation method. Subsequently, with such a dataset, we first fine-tune the base model of LLaMA-3 and LLaMA-3.1 to develop the code review model of \ECR\ and then align \ECR\ to develop an enhanced model of \AECR.

\begin{figure*}[ht]
    \centering
    \includegraphics[width=7in]{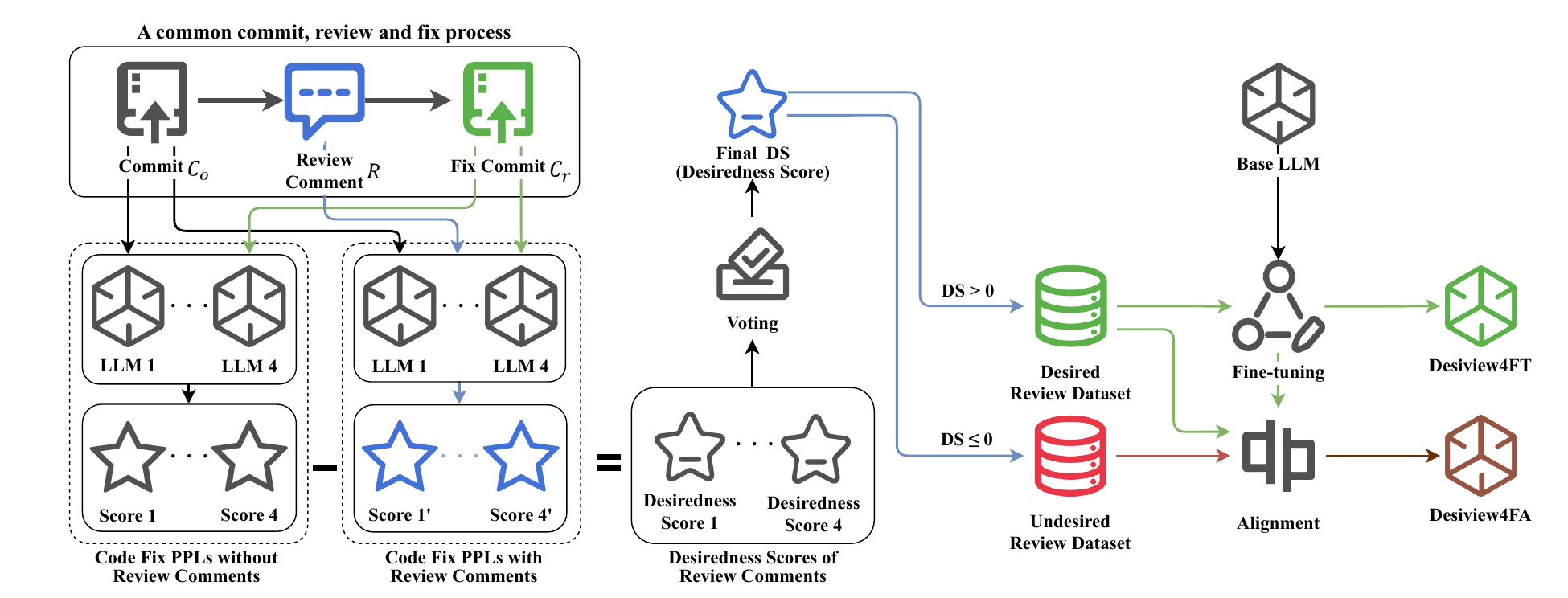}
    \caption{The process of developing \ECR\ and \AECR}
    \label{fig:overview}
\end{figure*}

\subsection{\DSD: Constructing a distilled dataset}\label{sec:dsd}
The proposed \DSD\ dataset distillation method comprises two main steps: (1) identification of \DR{s}, and (2) dataset preparation and pre-processing.

\subsubsection{Identification of \DR{s}}
In theory, during a generation process, an LLM gradually generates content by continuously sampling data from the probability distribution of the next token, in which
tokens with higher probabilities are more likely to be selected.
When the average probability of the given answer is higher, the model is considered to be more certain about that answer. Based on this principle, researchers \cite{jelinek1977perplexity, miaschi2021makes} have proposed the concept of `perplexity' and used it to evaluate models and guide the selection of hyperparameters \cite{miaschi2021makes}. Generally, the definition of `perplexity' is as follows:
\begin{equation}
\mathbf{PPL}(X) = exp\{-\frac{1}{N}\sum^N_{i=1}logP(x_i|x_{<i})\}
\end{equation}
where $X= (x_0,x_1,...,x_N)$ is the answer to be evaluated, $x_i$ is the \(i\)-th token, $logP(x_i|x_{<i})$ is the log-likelihood of the \(i\)-th token given the preceding tokens, and \(N\) is the total number of tokens to be calculated. Perplexity is used to evaluate the model's ability to uniformly predict a specified set of tokens for a given content. The higher the perplexity, the lower the probability that the model successfully generates the given content, and vice versa.

For a code review task, the reviewer first writes review comments $R$ based on the original code commit $C_o$, denoted by $P(R|C_o)$. Subsequently, the developer writes code fixes $C_r$ based on the review comments $R$, denoted by $P(C_r|C_o,R)$, as shown in the upper left of Fig. \ref{fig:overview}. Since \DR{s} should lead to code fixes, as pointed out in several studies \cite{kononenko2016code}, we can calculate the \textit{desiredness score} of review comments \(DS\) according to the following formula:

\begin{equation}
DS = -(\mathbf{PPL}(P(C_r|C_o,R))-\mathbf{PPL}(P(C_r|C_o)))
\label{desiredScore}
\end{equation}
The formula represents the difference in the perplexity of the code fix with and without the review comments, using a negative sign to align with the human preference that higher scores indicate more desired comments. Generally, when \(DS > 0\), it is considered that the review comments have had a positive impact on the code fix, making them desired. When \(DS \leq 0\), it is considered that the review comments have not contributed to the code fix or have introduced noises, making them undesired.

\subsubsection{Dataset preparation and pre-processing}

We select the CodeReviewer dataset \cite{li2022codereviewer}, one of the most widely adopted datasets in code review research, as the base dataset to construct a distilled dataset. To the best of our knowledge, this is the only public \M{multi-programming language} dataset in code review research field that contains the original code submissions ($C_o$), code review comments ($R$), and subsequent code fixes ($C_r$), thereby meeting all the requirements for identifying \DR{s} elaborated above. To perform the perplexity calculation, we use a straightforward prompt, as shown in Fig. \ref{fig:CodeRefineTemplate}, to generate code fixes with and without the review comments. The desiredness score of \DR{s} is then calculated as the difference between the perplexity for both code fixes with and without the review comments. An example of perplexity calculation is shown in Fig. \ref{fig:EffectivenessExample}. We construct the dialogue input using the chat templates of different models and calculate the perplexity of the standard answers to obtain the required perplexity. Note that the perplexity calculation does not involve a content generation process, so it is not affected by errors due to LLM hallucinations. When this score is greater than 0, the review comment is judged to be desired; otherwise, it is considered undesired.

\begin{figure}[bpht!]
    \centering
    \scriptsize
    \begin{lstlisting}[frame=single]
Refine the given code based on the provided code review 
comment.
The comment is: '{comment}'
The code is: '{code}'
    \end{lstlisting}
    \caption{Code refine template}
    \label{fig:CodeRefineTemplate}
\end{figure}

\begin{figure*}[h]
    \centering
    \includegraphics[width=6.8in]{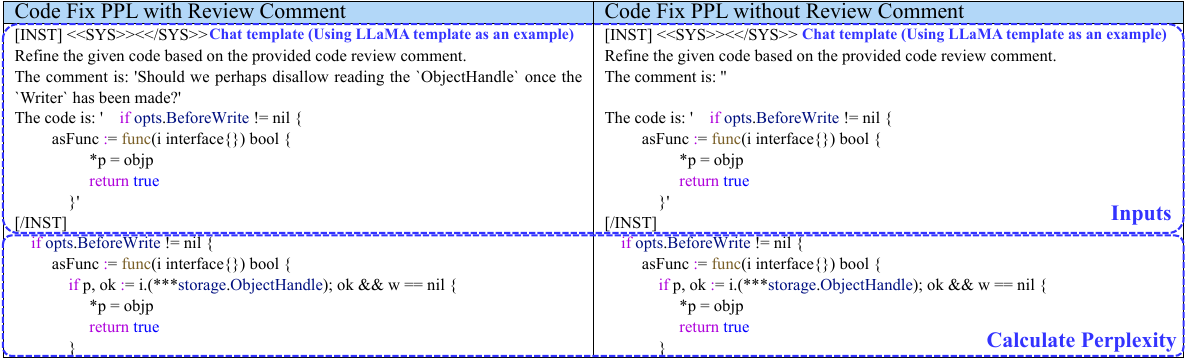}
    \caption{A perplexity calculation example}
    \label{fig:EffectivenessExample}
\end{figure*}

We use four commonly available LLMs to construct a consensus mechanism(i.e., voting) so as to enhance the accuracy of the resultant judgment, including: CodeLlama-13B \cite{roziere2023code}, starchat2-15B \cite{lozhkov2024starcoder}, Meta-Llama-3-8B \cite{llama3modelcard}, and deepseek-coder-6.7B \cite{guo2024deepseek}. The median of the results from the four LLMs is used as the final score to determine the desired review comments.
The identification results of the desired review comments of the training set and testing set in CodeReviewer dataset are shown in Table \ref{tab:DatasetResult}. We can observe that less than half of the review comments are \DR{s}, which have a positive effect on subsequent code fixes.
\M{In addition, the proportions of \DR{s} in the training and test sets are also close to each other, somewhat indicating the reliability of \DSD.}
\begin{table}[htbp]
\caption{Analysis Results of CodeReviewer Dataset}
\begin{center}
\scriptsize
\begin{tabular}{|c|c|c|c|}
\hline
Dataset Type & Total & Desired & Undesired \\
\hline
Training & 150406 (100\%) & 64934 (43.17\%) & 85472 (56.83\%) \\
\hline
Testing & 13103 (100\%) & 5727 (43.71\%) & 7376 (56.29\%) \\
\hline
\end{tabular}
\end{center}
\label{tab:DatasetResult}
\end{table}

\subsection{\ECR: Fine-tuning Large Language Models}

With the distilled dataset, we fine-tune base LLMs to develop our first code review model \ECR. We choose LLaMA series as the base LLMs since they are among the most commonly used models in the open-source community \cite{zhao2023survey}. To be specific, we use both LLaMA-3 \cite{llama3modelcard} and the most recently released LLaMA-3.1 since both LLMs represent the latest models in the LLaMA series. In particular, we use the smallest version of these models, namely LLaMA-3-8B and LLaMA-3.1-8B, due to our GPU resource limitations.

In terms of training methods, we opt to use LoRA \cite{hu2021lora} for fine-tuning the LLaMA series, thereby reducing the resource requirements. LoRA assumes that the parameter changes during the fine-tuning phase have a low intrinsic rank, allowing the parameter changes to be decomposed into the product of low-rank matrices, i.e., $W^{\prime}=W_0+\Delta W=W_0+BA$. Here, $W^{\prime}$ represents the fine-tuned model parameters, $W_0$ is the set of pre-trained model parameters, $\Delta W$ is the change in model parameters after fine-tuning, $B \in \mathbb{R}^{d \times r}$, $A \in \mathbb{R}^{r \times k}$, with $d$ and $k$ being the dimensions of the model parameters, and satisfying $r \ll \min(d,k)$. During training, the original pre-trained parameter set $W_0$ is frozen and does not participate in gradient updates; only $B$ and $A$ are updated. Since the number of parameters in the low-rank matrices is much smaller than that of the original model matrix, it allows for fine-tuning the large model with a minimal number of parameters. The fine-tuning was conducted using 2 Nvidia A100 40GB GPUs, with the fine-tuning parameters shown in Table \ref{tab:TrainingHyp}. The prompts used for fine-tuning were based on the LLaMA-Reviewer prompts to facilitate subsequent comparisons, and the code review task is illustrated in Fig. \ref{fig:CodeReviewTemplate}.
\begin{table}[htbp]
\caption{Training hyperparameters}
\begin{center}
\scriptsize
\begin{tabular}{|c|c|c|c|c|c|c|c|}
\hline
Method & epochs & batch & lr & cutoff & lora r & \makecell[c]{lora\\alpha} & \makecell[c]{lora\\dropout} \\
\hline
Fine-tuning & 10 & 32 & 1e-5 & 2048 & 16 & 32 & 0.05 \\
\hline
Alignment & 5 & 64 & 1e-5 & 2048 & 16 & 32 & 0.05 \\
\hline
\end{tabular}
\end{center}
\label{tab:TrainingHyp}
\end{table}
\begin{figure}[bpht!]
    \centering
    \scriptsize
    \begin{lstlisting}[frame=single]
Review the given code and provide a constructive code 
review comment.
The code/(diff hunk) is: '{} '
    \end{lstlisting}
    \caption{Code Review template}
    \label{fig:CodeReviewTemplate}
\end{figure}
\subsection{\AECR: Aligning Large Language Models}
While LLM fine-tuning with task-specific data can improve its task performance, LLM alignment goes a step further by ensuring the LLM behaves in accordance with human intentions and values. Therefore, we align model \ECR\ to develop an enhanced code review model \AECR\ by encouraging LLMs to generate desired review comments. 
LLM alignment typically requires paired data, i.e., a desired answer and an undesired answer under the same prompt. However, in code review, there can be usually only one review comment within a piece of code, making it difficult to construct reasonable paired data. Therefore, we choose the KTO algorithm \cite{ethayarajh2024kto}, which does not require paired data. The optimization objective of KTO is as follows:

$$
L_{KTO}(\pi_\theta,\pi_{ref}) = \mathbb{E}_{x,y \sim D}[\lambda_y - v(x,y)]
$$
where:
$$
v(x,y) = \begin{cases}
\lambda_D \sigma(\beta(r_\theta(x,y) - z_0)) & \text{if } y \sim y_{desired}|x \\
\lambda_U \sigma(\beta(z_0 - r_\theta(x,y))) & \text{if } y \sim y_{undesired}|x
\end{cases}
$$
$$
r_\theta(x,y) = \log \frac{\pi_\theta(y|x)}{\pi_{ref}(y|x)}
$$
$$
z_0 = \mathbb{E}_{x' \sim D}[\mathbf{KL} (\pi_\theta(y'|x')||\pi_{ref}(y'|x'))]
$$
$$
\frac{\lambda_D n_D}{\lambda_U n_U} \in [1, \frac{4}{3}]
$$
$\pi_\theta$ is the model to be optimized, which in this work is the fine-tuned model with the LoRA model superimposed, where LoRA is the trainable part. $\pi_{ref}$ is the reference model, which in this work is the fine-tuned model. A \textbf{KL} (Kullback–Leibler) divergence penalty is introduced to restrict how far the language model can drift from $\pi_{ref}$. $\lambda_y$ is usually set to 1, and $\lambda_D$ and $\lambda_U$ are set according to the ratio of desirable data $n_D$ and undesirable data $n_U$ and the constraint $\frac{\lambda_D n_D}{\lambda_U n_U} \in [1, \frac{4}{3}]$, with $\lambda_D = 1.7$ and $\lambda_U = 1.0$. $\sigma$ is a nonlinear function, here taken as sigmoid, and $\beta$ is used to control the degree of risk aversion. The larger the value, the more quickly the value saturates, meaning the model is simultaneously more risk-averse in gains and more risk-seeking in losses. This value is set to 0.1, consistent with the original paper~\cite{ethayarajh2024kto}. To reduce the GPU resource requirements for training, alignment also uses LoRA for training. Other training hyperparameters are also shown in Table \ref{tab:TrainingHyp}.

\section{Evaluation}\label{sec:eval}
In this section, we validate the performance of the \DSD\ dataset distillation method for identifying \DR{s} and examine the effect of using the distilled dataset to fine-tune and align LLMs on their ability to perform code review tasks. Specifically, we aim to answer the following research questions:
\begin{itemize}
\item RQ1: How accurately can the dataset distillation method identify \DR{s}? 
\item RQ2: How much performance enhancement can LLMs gain by being fine-tuned and aligned with the distilled dataset?
\end{itemize}

RQ1 aims to gauge the effectiveness of the proposed \DSD\ dataset distillation method in identifying desired review comments and subsequently constructing a high-quality distilled dataset compared to that of existing alternative methods. RQ2 aims to test the hypothesis that LLMs fine-tuned and aligned with the distilled dataset (specifically \ECR\ and \AECR) can generate more desired review comments than those fine-tuned and aligned with the original dataset (specifically LLaMA-Reviewer).

\begin{figure}[]
    \centering
    \includegraphics[width=3.4in]{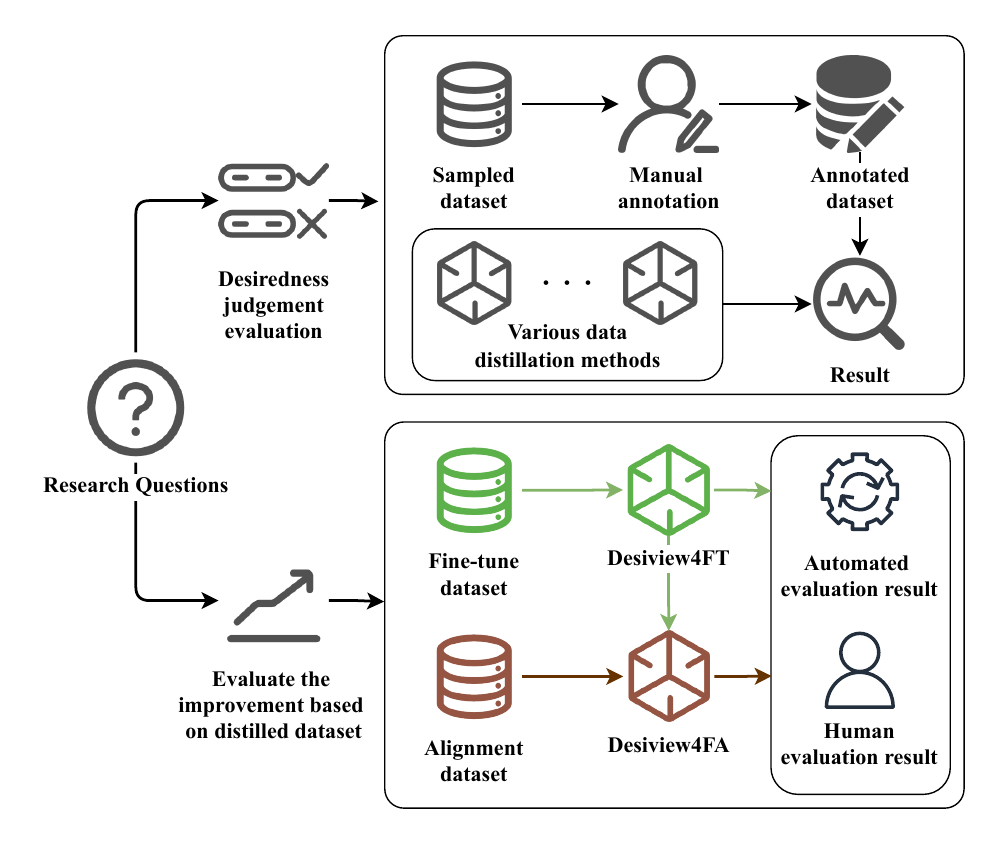}
    \caption{The evaluation process}
    \label{fig:evaluate}
\end{figure}

\subsection{Experimental settings}

\underline{Dataset.} The base dataset is CodeReviewer dataset~\cite{li2022codereviewer}, which is the only public multi-programming language dataset for code review research in the open-source community and has been widely used in several studies \cite{li2022codereviewer, lu2023llama}. Besides, as pointed out in Section~\ref{sec:dsd}, it is so far the only publicly available dataset that contains code snippets before and after the review as well as the review comments that meet the needs of this study.

\underline{Benchmark approaches}.
For RQ1, we aim to compare the effectiveness of different methods for identifying \DR{s}. We choose the 10-line rule \cite{bosu2015characteristics}, GPT-3.5, and GPT-4o as the benchmark methods. The first method is one of the few rule-based approaches and has been adopted by several studies \cite{bosu2015characteristics, rong2024distilling}. Meanwhile, the GPT family has been widely used in numerous studies as a benchmark method for text comprehension and analysis. The latter, in particular, has been confirmed by many studies as one of the strongest LLMs available for accomplishing such tasks. For RQ2, we select LLaMA-Reviewer \cite{lu2023llama} as the benchmark method because it uses the same dataset, and, additionally, our study also uses LLaMA as the base model. Choosing LLaMA-Reviewer as the baseline approach facilitates a fair comparison.

\subsubsection{The experiment for RQ1}
First of all, we need to construct a test set containing explicit annotations of \DR\ and non-\DR\ for each entry.
As shown in Table \ref{tab:DatasetResult}, the CodeReviewer training set contains a total of 150,406 entries. We randomly selected 600 of these entries for manual annotation, achieving a margin of error of less than 4\% at a 95\% confidence level. 
The annotation was performed by two software engineering graduate students\M{, each annotating 450 data entries with 300 duplicated entries to check for consistency. To be specific, when a review comment triggers a fix that pertains to the review comment, the review comment is labeled ``\textit{desired}." Otherwise, it is labeled ``\textit{undesired}."} 
We used the Chi-Squared test \cite{pearson1900x} to check the determination consistency of the \M{duplicate annotations} and obtained a p-value of 0.965, thereby rejecting the hypothesis of inconsistency in the annotations.

To demonstrate the effectiveness of the \DSD\ dataset distillation method, we compare it against the 10-line rule for change-triggering review comments \cite{bosu2015characteristics}, GPT-3.5-turbo, and GPT-4o using prompt engineering. Different treatments are required for these benchmark approaches.

\underline{The 10-line rule} determines whether changes were made within 10 lines of the given review comment. As the CodeReviewer dataset only contains information on changes at the corresponding locations, the rule was simplified to whether modifications were made subsequently.

\underline{GPT-3.5-turbo and GPT-4o} require prompt engineering to detect \DR{s}, as shown in Fig. \ref{fig:APrompt}. We experimented with different phrasing methods and selected a relatively better-performing prompt as the final prompt. To avoid the impact of sampling generation by large models, we set the sampling parameter temperature to 0, ensuring the model uses greedy search generation for result stability. As other methods did not provide examples, to ensure a fair comparison, we did not use examples in the prompt engineering method either, i.e., we adopted a zero-shot prompt \cite{brown2020language} strategy. Common metrics such as `accuracy', `precision', `recall', and `F1-score' were used for evaluation, thereby determining the effectiveness of different methods in identifying \DR(s).

\begin{figure}[bpht!]
    \centering
    \scriptsize
    \begin{lstlisting}[frame=single]
Your task is to determine whether the changes in the given 
original code and the modified code pertain to the provided
review comment. If they pertain, output True; if they do not
pertain, output False. Only provide True or False, without 
any additional content.
```original code
{}
```
```modified code
{}
```
```review comment
{}
```

    \end{lstlisting}
    \caption{The prompt used to detect \DR{s}}
    \label{fig:APrompt}
\end{figure}

\subsubsection{The experiment for RQ2}
To evaluate the quality of \DR{s} generated by LLMs trained with the original and the distilled datasets, we compare three LLMs: (1) LLaMA (LLaMA-3 and LLaMA-3.1) fine-tuned with the original dataset, i.e., LLaMA-Reviewer\cite{lu2023llama}, (2) LLaMA (LLaMA-3 and LLaMA-3.1) fine-tuned with the distilled data, i.e., \ECR, and LLaMA (LLaMA-3 and LLaMA-3.1) aligned with both distilled data (\DR{s}) and dropped data (non-\DR{s}), i.e., \AECR. For a fair comparison, the fine-tuning process applies the same settings as study~\cite{lu2023llama} with different datasets. The evaluation process consists of two parts: automated evaluation and human evaluation.

\underline{Automated evaluation} uses a test set of 5,727 entries, as shown in Table \ref{tab:DatasetResult}. \M{As the distilled dataset contain a high proportion of review comments that can lead to effective code fixes, it is fair enough to regard the ground truth as the correct answer.} With the trained LLMs generating review comments for a given code commit, the generated review comments are compared against the existing ones contained in the test set using the BLEU-4 metric \cite{papineni2002bleu} to calculate text similarity.

\underline{Human evaluation} is conducted by two software engineering graduate students. The test set contains a total of 5,727 \DR\ entries (as shown in Table \ref{tab:DatasetResult}), from which we randomly selected 300 entries for human evaluation, ensuring a margin of error of less than 6\% at a 95\% confidence level. \M{Each student evaluates 180 pieces of data, including 60 duplicated evaluations, to check for consistency.} We applied the Chi-Squared test \cite{pearson1900x} to check consistency, obtaining a p-value of 0.887, thereby rejecting the hypothesis of evaluation inconsistency and proving the consistency of the evaluations. The human evaluation involved observing the original code commit under review and the LLM-generated review comments to determine whether the provided review comments correctly identify and describe the issues. In this sense, we divided the evaluation into two tasks: accurately locating code issues and accurately describing the issues. The criteria we adopted to determine the results of these two tasks are as follows:
\begin{enumerate}
    \item \textit{Human Position:} To be considered correct, it requires the LLM-generated review comments to pinpoint the same location of the code issues as in the answer, regardless of whether the description of the code issues is correct or not. 
    \item \textit{Human Perfect:} To be considered correct, it requires the LLM-generated review comments to describe the same issues and/or solutions as in the answer. It is clear that the second task builds upon the first one.
\end{enumerate}

\subsection{Results analysis}
\begin{figure*}[ht]
    \centering
    \includegraphics[width=7in]{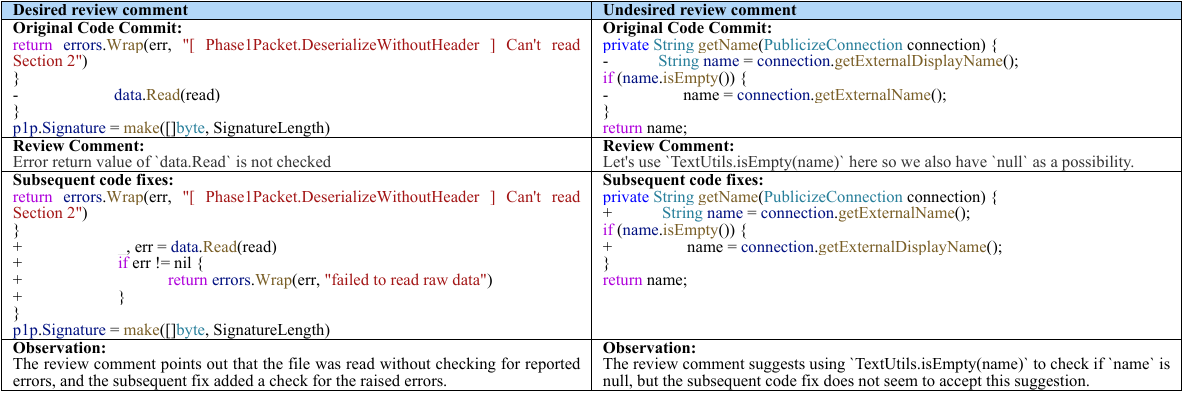}
    \caption{Examples of \DR{s} and non-\DR{s} identified by \DSD}
    \label{fig:Desirable_Undesirable}
\end{figure*}

\paragraph{RQ1: Accuracy of automated identification of \DR{s}}
\begin{table}[htbp]
\caption{Performance of each method in identifying \DR{s}}
\begin{center}
\scriptsize
\begin{tabular}{|c|c|c|c|c|}
\hline
Method & Accuracy & Precision & Recall & F1-Score  \\
\hline
10-line rule & 58.33 & 51.92 & \textbf{100.00} & 68.35  \\
\hline
gpt3.5-turbo-0125 & 68.00 & 60.71 & 81.85 & 69.72  \\
\hline
gpt-4o-0513 & 76.50 & 79.72 & 64.07 & 71.05  \\
\hline
\DSD & \textbf{86.67} & \textbf{88.93} & 80.37 & \textbf{84.44}  \\
\hline
\end{tabular}
\end{center}
\label{tab:APerformance}
\end{table}
Table \ref{tab:APerformance} presents the performance of different methods in identifying \DR{s}. As all entries in the test set contain code changes, the 10-line rule can always identify all \DR{s}, achieving 100\% recall. However, this method only determines changes and cannot assess whether the changes align with the review comments, resulting in poor performance in comprehensive metrics such as accuracy and F1-score. The GPT-3.5-turbo and GPT-4o methods can somewhat understand the relationship between changes and review comments, but their performance is still inferior to our method, which significantly outperforms existing methods in both comprehensive metrics of accuracy and F1-score. Fig. \ref{fig:Desirable_Undesirable} illustrates examples of desired and undesired review comments. From these examples, it is evident that the determination of \DR{s} cannot be based solely on the review's phrasing or keywords to ascertain whether they can lead to effective code fixes. It requires a deeper understanding of both the code and the review comments.
\paragraph{RQ2: Effect of dataset distillation on the task performance of LLMs}

\begin{table}[htbp]
\caption{Performance of the code review comment generation task}
\begin{center}
\scriptsize
\begin{tabular}{|c|c|c|c|}
\hline
Method & BLEU-4 & Human Position & Human Perfect \\
\hline
\makecell[c]{LLaMA-Reviewer\\(LLaMA-3 Origin)} & 8.33 & 70.33 & 16.67  \\
\hline
\makecell[c]{\ECR\\(LLaMA-3 Based)} & \makecell[c]{11.87\\(+42.50\%)} & \makecell[c]{76.67\\(+9.01\%)} & \makecell[c]{18.33\\(+9.96\%)} \\
\hline
\makecell[c]{\AECR\\(LLaMA-3 Based)} & \makecell[c]{13.13\\(+57.62\%)} & \makecell[c]{\textbf{80.00}\\(+13.75\%)} & \makecell[c]{\textbf{18.67}\\(+12.00\%)}  \\
\hline
\makecell[c]{LLaMA-Reviewer\\(LLaMA-3.1 Origin)} & 6.86 & 68.67 & 12.67  \\
\hline
\makecell[c]{\ECR\\(LLaMA-3.1 Based)} & \makecell[c]{12.48\\(+81.92\%)} & \makecell[c]{78.67\\(+14.56\%)} & \makecell[c]{16.00\\(+26.28\%)} \\
\hline
\makecell[c]{\AECR\\(LLaMA-3.1 Based)} & \textbf{\makecell[c]{13.57\\(+97.81\%)}} & \makecell[c]{79.00\\\textbf{(+15.04\%)}} & \makecell[c]{16.67\\\textbf{(+31.57\%)}}  \\
\hline
\end{tabular}
\end{center}
\label{tab:GenPerformance}
\end{table}

Table \ref{tab:GenPerformance} shows the performance of the code review comment generation task under both automated and human evaluation. It is evident that the distilled dataset significantly improves the performance of the LLMs in generating \DR{s} that present more accurate and useful information to users. Notably, LLMs fine-tuned with the distilled dataset that contains a high proportion of \DR{s} and whose size is less than half of that of the original dataset significantly outperform those fine-tuned with the original full dataset in terms of both localizing and describing code issues. Alignment further extends this advantage. It is worth noting that LLaMA-3.1 does not show a clear advantage over LLaMA-3, \M{but it appears to be more sensitive to the fine-tuning and alignment of the distilled dataset}. Nevertheless, the distilled dataset improves both versions.

\begin{figure*}[!ht]
    \centering
    \includegraphics[width=7in]{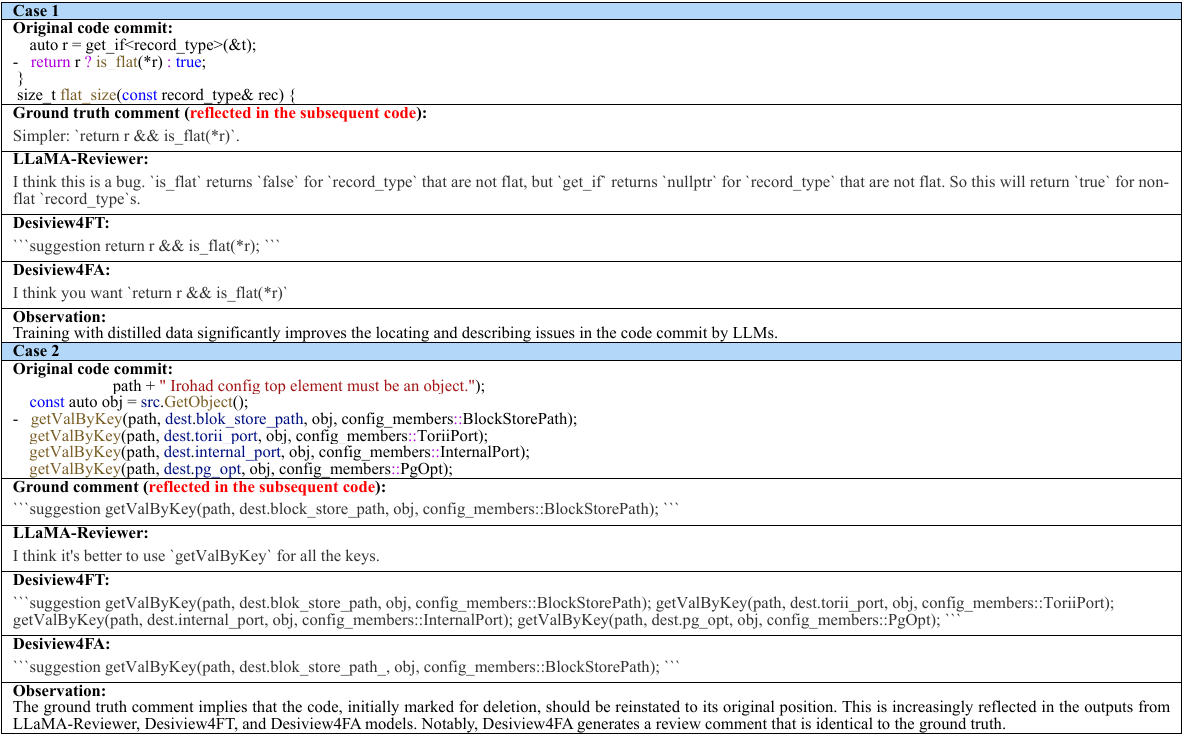}
    \caption{Examples of review comments generated by LLaMA-Reviewer, \ECR\ and \AECR}
    \label{fig:GenExample}
\end{figure*}

Fig. \ref{fig:GenExample} presents some intuitive examples of the review comments generated by the three methods (based on LLaMA-3 only). It is apparent that training with the distilled dataset significantly enhances LLaMA’s ability to identify key issues. Using alignment techniques can further improve the model’s ability to generate accurate information and reduce the occurrence of irrelevant information.

\section{Discussion}\label{sec:discussion}
The primary contribution of this work is the dataset distillation method, \DSD, which can be used to construct a distilled dataset for fine-tuning LLMs to enhance their performance in code review tasks. This contribution has a profound impact on LLM-based code review research and can be generally extended to other LLM-based software engineering tasks. In this section, we discuss the implications of dataset distillation in code review research.

\subsection{Distilled dataset for training code review models}
Training data is fundamental to generating effective automated code reviews. The results of our study partially demonstrate the value a distilled dataset brings to LLM-based code review, confirming that distilled data often leads to better LLM performance \cite{zhou2024lima, wei2024magicoder}. However, acquiring distilled datasets is generally challenging. Specifically, in the field of code review, previous studies have not effectively addressed this issue, aside from the extremely costly manual annotation methods \cite{ouyang2022training, mcaleese2024llm}. By leveraging the relationship between the content of review comments and the code fixes as a criterion, we introduce the perplexity metric to the quality assessment of code review data in terms of the desiredness of review comments. The proposed dataset distillation method enables automated and reliable acquisition of large-scale, high-quality review data at a low cost, thereby effectively addressing the scarcity of high-quality code review datasets. More importantly, \M{we believe this method has the potential to be customized and generalized to support other code review objectives, such as performance bottlenecks~\cite{shypula2023learning} and security risks~\cite{tihanyi2023formai} and even other software engineering objectives, such as vulnerability detection~\cite{croft2023data}.  Fig. \ref{fig:overview} (upper left corner) illustrates a typical pull-request development mode widely used in the open-source community. Drawing on the principles behind \DSD, it should be feasible to construct high-quality datasets applicable to various LLM-enabled software engineering scenarios by adjusting the desiredness criterion of review comments to other relevant criteria, which necessitates further exploration.}

\subsection{Distilled dataset for training review quality prediction models}
The distilled dataset obtained from this study can also be used to train models to predict the quality of generated review comments — specifically, predicting whether they can trigger code fixes. One of the pain points in applying LLMs to automated code review is the uncontrollable quality of generated review comments. In extreme cases, developers have to check not only the code but also the generated review comments, which can even increase their workload. This defeats the purpose of automated code review, which is to reduce their workload. By applying the proposed dataset distillation method, which divides the original dataset into high-quality and low-quality datasets, we can train a binary classification model on traditional Bert series~\cite{liu2019roberta, feng2020codebert} models with smaller parameter sizes to predict the quality of LLM-generated review comments. The training hyperparameters are shown in Table \ref{tab:DesirablePHyp}.

\begin{table}[htbp]
\caption{Training hyperparameters for predicting desired review comments }
\begin{center}
\scriptsize
\begin{tabular}{|c|c|c|c|c|c|}
\hline
Pattern & epochs & batch & lr & label smoothing & weight decay  \\
\hline
Value & 5 & 32 & 1e-5 & 0.1 & 0  \\
\hline
\end{tabular}
\end{center}
\label{tab:DesirablePHyp}
\end{table}

\begin{table}[htbp]
\caption{Performance of the \DR\  prediction task}
\begin{center}
\scriptsize
\begin{tabular}{|c|c|c|c|c|}
\hline
Method & Accuracy & Precision & Recall & F1-Score  \\
\hline
Robert-Base & 79.82 & 84.28 & 80.71 & 82.46  \\
\hline
CodeBert-base & 79.11 & 86.82 & 78.39 & 82.39  \\
\hline
\end{tabular}
\end{center}
\label{tab:RCPerformance}
\end{table}

The review comment quality prediction task involves predicting whether a given review comment will lead to a modification based on the original code submission and the generated review comment. The input consists of a code Diff from the original submission and a corresponding review comment, and the output indicates whether it will lead to a code fix. We selected two commonly used Bert series models, Roberta \cite{liu2019roberta} and CodeBERT \cite{feng2020codebert}, as our base models. Evaluation metrics include accuracy, precision, recall, and F1-score. The results in Table \ref{tab:RCPerformance} show a great potential to predict the quality of generated review comments. 
One possible application scenario is to integrate an LLM-based code review solution with a review comment quality assessment mechanism to provide a corresponding quality evaluation for each review comment, which can assist developers in making better decisions on whether to accept a particular review comment. A more aggressive approach could be to embed this quality assessment mechanism within the LLM-based code review system, ensuring it only outputs high-quality review comments and drops low-quality ones.

\section{Threats to validity}\label{sec:threat}
In this section, we discuss several validity risks. 

\paragraph*{Definition of quality in terms of desiredness}
In this paper, we define a high-quality code review dataset as one composed of only \DR{s}. This approach inevitably carries some validity risks. The concept of \DR{s} in this paper specifically refers to review comments that can trigger subsequent code fixes and improvements. This concept is based on multiple code review studies, all of which regard the detection of issues and triggering of subsequent code fixes as their primary purpose \cite{kononenko2016code}. Our study does not intend to diminish other purposes or the unique meaning of \DR{s} in these scenarios. For instance, as shown in Fig. 1 (example C), acknowledging a developer's fix can also be meaningful to the developer. In fact, during our exploration process, we found that the high-quality dataset distilled according to our definition of \DR\ is large enough (close to 65,000 entries of \DR{s}) to meet the needs for fine-tuning LLaMA. Therefore, we did not expand the concept of \DR{s} to cover other code review purposes. Nevertheless, this leaves some room for future research.

\M{
\paragraph*{Noise in the distilled dataset}
Based on the results in Table \ref{tab:APerformance}, it is reasonable to assume that a small number of non-\DR{s} remain in the distilled dataset when using \DSD\ to distil the CodeReviewer dataset, including the test set. This may introduce some bias in addressing RQ2. However, the amount of such data is minimal and unlikely to have a significant impact. Moreover, the results of the human evaluation fully corroborate the validity and efficacy of the \DSD\ method, making this risk controllable.
}

\paragraph*{Pre-trained models}
Different pre-trained models exhibit varying performance in the task of code review. During the phase of assessing \DR{s}, we employed a voting mechanism using multiple models (without any preference) to reduce the impact of a single LLM on the results. Although the results show a positive effect of this strategy, it is possible that using different LLMs may impact the results. 

\paragraph*{Model parameter size}

In general, the performance of a model can be significantly influenced by the size of its parameters within the same series. However, due to limitations in GPU capacity, we restricted our study to models with parameters under 16B for all computations. For tasks such as fine-tuning and alignment, we utilized models of 8B size. It is worth noting that employing larger models could potentially enhance performance, particularly in terms of assessing the quality of review comments and generating desired review comments more accurately.
The size of the model parameters also contributes to a substantial disparity in the quality of the review comments we generate compared to those generated by GPT4o. We believe one of the reasons for this difference is on the scale of three orders of magnitude in terms of parameter numbers. Despite this, our distilled dataset has proven highly effective in enhancing the original LLaMA series. Therefore, it would be intriguing to explore and compare the effectiveness of LLaMA-3 and GPT4o in conducting code review with a comparable number of parameters (e.g., LLaMA-3.1-405B).

\paragraph*{Model training methods}
Typically, large model training methods are divided into low-parameter training and full-parameter training. Low-parameter training involves fine-tuning only a small portion of the LLM, making it possible to train it with fewer resources. Full-parameter training involves training all parameters of the LLM, which can lead to better performance but at a significantly higher cost compared to low-parameter training \cite{biderman2024lora}. Research has shown that full-parameter training outperforms low-parameter training. Due to computational resource constraints, this work employed low-parameter training methods for training LLMs. Using full-parameter training methods might result in better performance.

\paragraph*{Dataset}
As far as we know, the only public \M{multi-programming language} dataset in the open-source community that includes original submissions, subsequent fixes, and review comments is the CodeReviewer dataset\cite{li2022codereviewer}. Therefore, we could only use this dataset for our training and testing. This factor creates some risk in terms of the generalization of results, and we encourage the community to construct other datasets using our method to validate this work.

\paragraph*{Errors from human evaluation}
Despite using standardized methods and consistency checks to ensure the accuracy of manual evaluations, errors in manual evaluation are still possible, which may affect the results. \M{We mitigated this risk by employing evaluators with a background in software engineering, ensuring they have the necessary expertise to determine the relationship between source code and review comments during the evaluation. In addition, allowing some degree of duplication across multiple evaluators, in addition to conducting chi-square tests, further reduces this risk.}

\section{Conclusions}\label{sec:conclusion}
In this work, we propose a method for analyzing, assessing and automatically identifying \DR{s}. This solves one critical problem of automated construction of high-quality datasets for code review research. Empirical experiments reveal that this method surpasses all other methods in terms of identifying \DR{s}, including GPT-4o, in terms of accuracy. Using this method, we constructed a distilled dataset containing a high proportion of \DR{s}, which not only can be used to train a model to predict whether a new review comment is \DR\  but also support fine-tuning and aligning LLMs to perform better code review tasks in terms of generating \DR{s}. Both automated evaluation and human evaluation reveal that LLMs trained with the distilled dataset outperform those trained with the original dataset. Future work includes applying the proposed dataset distillation method to construct datasets suitable for different code review objectives, during which better LLMs can be leveraged to improve the accuracy of high-quality review comment identification. Additionally, using newer and stronger LLMs as the base models, along with new techniques for fine-tuning and alignment can also be explored to further enhance the application efficacy of distilled datasets.

\bibliographystyle{IEEEtran}
\balance
\bibliography{Citations}

\begin{thebibliography}{10}
\providecommand{\url}[1]{#1}
\csname url@samestyle\endcsname
\providecommand{\newblock}{\relax}
\providecommand{\bibinfo}[2]{#2}
\providecommand{\BIBentrySTDinterwordspacing}{\spaceskip=0pt\relax}
\providecommand{\BIBentryALTinterwordstretchfactor}{4}
\providecommand{\BIBentryALTinterwordspacing}{\spaceskip=\fontdimen2\font plus
\BIBentryALTinterwordstretchfactor\fontdimen3\font minus \fontdimen4\font\relax}
\providecommand{\BIBforeignlanguage}[2]{{%
\expandafter\ifx\csname l@#1\endcsname\relax
\typeout{** WARNING: IEEEtran.bst: No hyphenation pattern has been}%
\typeout{** loaded for the language `#1'. Using the pattern for}%
\typeout{** the default language instead.}%
\else
\language=\csname l@#1\endcsname
\fi
#2}}
\providecommand{\BIBdecl}{\relax}
\BIBdecl

\bibitem{gousios2014exploratory}
G.~Gousios, M.~Pinzger, and A.~v. Deursen, ``An exploratory study of the pull-based software development model,'' in \emph{Proceedings of the 36th international conference on software engineering}, 2014, pp. 345--355.

\bibitem{lu2023llama}
J.~Lu, L.~Yu, X.~Li, L.~Yang, and C.~Zuo, ``Llama-reviewer: Advancing code review automation with large language models through parameter-efficient fine-tuning,'' in \emph{2023 IEEE 34th International Symposium on Software Reliability Engineering (ISSRE)}.\hskip 1em plus 0.5em minus 0.4em\relax IEEE, 2023, pp. 647--658.

\bibitem{kononenko2016code}
O.~Kononenko, O.~Baysal, and M.~W. Godfrey, ``Code review quality: How developers see it,'' in \emph{Proceedings of the 38th international conference on software engineering}, 2016, pp. 1028--1038.

\bibitem{bosu2013impact}
A.~Bosu and J.~C. Carver, ``Impact of peer code review on peer impression formation: A survey,'' in \emph{2013 ACM/IEEE International Symposium on Empirical Software Engineering and Measurement}.\hskip 1em plus 0.5em minus 0.4em\relax IEEE, 2013, pp. 133--142.

\bibitem{li2022codereviewer}
Z.~Li, S.~Lu, D.~Guo, N.~Duan, S.~Jannu, G.~Jenks, D.~Majumder, J.~Green, A.~Svyatkovskiy, S.~Fu \emph{et~al.}, ``Automating code review activities by large-scale pre-training,'' in \emph{Proceedings of the 30th ACM Joint European Software Engineering Conference and Symposium on the Foundations of Software Engineering}, 2022, pp. 1035--1047.

\bibitem{hou2023large}
X.~Hou, Y.~Zhao, Y.~Liu, Z.~Yang, K.~Wang, L.~Li, X.~Luo, D.~Lo, J.~Grundy, and H.~Wang, ``Large language models for software engineering: A systematic literature review,'' \emph{arXiv preprint arXiv:2308.10620}, 2023.

\bibitem{fan2023large}
A.~Fan, B.~Gokkaya, M.~Harman, M.~Lyubarskiy, S.~Sengupta, S.~Yoo, and J.~M. Zhang, ``Large language models for software engineering: Survey and open problems,'' in \emph{2023 IEEE/ACM International Conference on Software Engineering: Future of Software Engineering (ICSE-FoSE)}.\hskip 1em plus 0.5em minus 0.4em\relax IEEE, 2023, pp. 31--53.

\bibitem{zhou2024lima}
C.~Zhou, P.~Liu, P.~Xu, S.~Iyer, J.~Sun, Y.~Mao, X.~Ma, A.~Efrat, P.~Yu, L.~Yu \emph{et~al.}, ``Lima: Less is more for alignment,'' \emph{Advances in Neural Information Processing Systems}, vol.~36, 2024.

\bibitem{liu2024coachlm}
Y.~Liu, S.~Tao, X.~Zhao, M.~Zhu, W.~Ma, J.~Zhu, C.~Su, Y.~Hou, M.~Zhang, M.~Zhang \emph{et~al.}, ``Coachlm: Automatic instruction revisions improve the data quality in llm instruction tuning,'' in \emph{2024 IEEE 40th International Conference on Data Engineering (ICDE)}.\hskip 1em plus 0.5em minus 0.4em\relax IEEE, 2024, pp. 5184--5197.

\bibitem{rejeleene2024towards}
R.~Rejeleene, X.~Xu, and J.~Talburt, ``Towards trustable language models: Investigating information quality of large language models,'' \emph{arXiv preprint arXiv:2401.13086}, 2024.

\bibitem{ouyang2022training}
L.~Ouyang, J.~Wu, X.~Jiang, D.~Almeida, C.~Wainwright, P.~Mishkin, C.~Zhang, S.~Agarwal, K.~Slama, A.~Ray \emph{et~al.}, ``Training language models to follow instructions with human feedback,'' \emph{Advances in neural information processing systems}, vol.~35, pp. 27\,730--27\,744, 2022.

\bibitem{mcaleese2024llm}
N.~McAleese, R.~M. Pokorny, J.~F.~C. Uribe, E.~Nitishinskaya, M.~Trebacz, and J.~Leike, ``Llm critics help catch llm bugs,'' \emph{arXiv preprint arXiv:2407.00215}, 2024.

\bibitem{wang2022self}
Y.~Wang, Y.~Kordi, S.~Mishra, A.~Liu, N.~A. Smith, D.~Khashabi, and H.~Hajishirzi, ``Self-instruct: Aligning language models with self-generated instructions,'' \emph{arXiv preprint arXiv:2212.10560}, 2022.

\bibitem{zhao2023survey}
W.~X. Zhao, K.~Zhou, J.~Li, T.~Tang, X.~Wang, Y.~Hou, Y.~Min, B.~Zhang, J.~Zhang, Z.~Dong \emph{et~al.}, ``A survey of large language models,'' \emph{arXiv preprint arXiv:2303.18223}, 2023.

\bibitem{luo2024semi}
X.~Luo, Q.~Zhu, Z.~Zhang, X.~Wang, Q.~Yang, D.~Xu, and W.~Che, ``Semi-instruct: Bridging natural-instruct and self-instruct for code large language models,'' \emph{arXiv preprint arXiv:2403.00338}, 2024.

\bibitem{plank2022problem}
B.~Plank, ``The'problem'of human label variation: On ground truth in data, modeling and evaluation,'' \emph{arXiv preprint arXiv:2211.02570}, 2022.

\bibitem{bosu2015characteristics}
A.~Bosu, M.~Greiler, and C.~Bird, ``Characteristics of useful code reviews: An empirical study at microsoft,'' in \emph{2015 IEEE/ACM 12th Working Conference on Mining Software Repositories}.\hskip 1em plus 0.5em minus 0.4em\relax IEEE, 2015, pp. 146--156.

\bibitem{sadowski2018modern}
C.~Sadowski, E.~S{\"o}derberg, L.~Church, M.~Sipko, and A.~Bacchelli, ``Modern code review: a case study at google,'' in \emph{Proceedings of the 40th international conference on software engineering: Software engineering in practice}, 2018, pp. 181--190.

\bibitem{shi2019automatic}
S.-T. Shi, M.~Li, D.~Lo, F.~Thung, and X.~Huo, ``Automatic code review by learning the revision of source code,'' in \emph{Proceedings of the AAAI Conference on Artificial Intelligence}, vol.~33, no.~01, 2019, pp. 4910--4917.

\bibitem{feng2020codebert}
Z.~Feng, D.~Guo, D.~Tang, N.~Duan, X.~Feng, M.~Gong, L.~Shou, B.~Qin, T.~Liu, D.~Jiang \emph{et~al.}, ``Codebert: A pre-trained model for programming and natural languages,'' \emph{arXiv preprint arXiv:2002.08155}, 2020.

\bibitem{wang2021codet5}
Y.~Wang, W.~Wang, S.~Joty, and S.~C. Hoi, ``Codet5: Identifier-aware unified pre-trained encoder-decoder models for code understanding and generation,'' \emph{arXiv preprint arXiv:2109.00859}, 2021.

\bibitem{radford2019language}
A.~Radford, J.~Wu, R.~Child, D.~Luan, D.~Amodei, I.~Sutskever \emph{et~al.}, ``Language models are unsupervised multitask learners,'' \emph{OpenAI blog}, vol.~1, no.~8, p.~9, 2019.

\bibitem{hong2022commentfinder}
Y.~Hong, C.~Tantithamthavorn, P.~Thongtanunam, and A.~Aleti, ``Commentfinder: a simpler, faster, more accurate code review comments recommendation,'' in \emph{Proceedings of the 30th ACM joint European software engineering conference and symposium on the foundations of software engineering}, 2022, pp. 507--519.

\bibitem{gupta2018intelligent}
A.~Gupta and N.~Sundaresan, ``Intelligent code reviews using deep learning,'' in \emph{Proceedings of the 24th ACM SIGKDD International Conference on Knowledge Discovery and Data Mining (KDD’18) Deep Learning Day}, 2018.

\bibitem{roziere2023code}
B.~Roziere, J.~Gehring, F.~Gloeckle, S.~Sootla, I.~Gat, X.~E. Tan, Y.~Adi, J.~Liu, T.~Remez, J.~Rapin \emph{et~al.}, ``Code llama: Open foundation models for code,'' \emph{arXiv preprint arXiv:2308.12950}, 2023.

\bibitem{touvron2023llama}
H.~Touvron, L.~Martin, K.~Stone, P.~Albert, A.~Almahairi, Y.~Babaei, N.~Bashlykov, S.~Batra, P.~Bhargava, S.~Bhosale \emph{et~al.}, ``Llama 2: Open foundation and fine-tuned chat models,'' \emph{arXiv preprint arXiv:2307.09288}, 2023.

\bibitem{guo2024deepseek}
D.~Guo, Q.~Zhu, D.~Yang, Z.~Xie, K.~Dong, W.~Zhang, G.~Chen, X.~Bi, Y.~Wu, Y.~Li \emph{et~al.}, ``Deepseek-coder: When the large language model meets programming--the rise of code intelligence,'' \emph{arXiv preprint arXiv:2401.14196}, 2024.

\bibitem{lozhkov2024starcoder}
A.~Lozhkov, R.~Li, L.~B. Allal, F.~Cassano, J.~Lamy-Poirier, N.~Tazi, A.~Tang, D.~Pykhtar, J.~Liu, Y.~Wei \emph{et~al.}, ``Starcoder 2 and the stack v2: The next generation,'' \emph{arXiv preprint arXiv:2402.19173}, 2024.

\bibitem{llama3modelcard}
\BIBentryALTinterwordspacing
AI@Meta, ``Llama 3 model card,'' 2024. [Online]. Available: \url{https://github.com/meta-llama/llama3/blob/main/MODEL_CARD.md}
\BIBentrySTDinterwordspacing

\bibitem{chen2022codet}
B.~Chen, F.~Zhang, A.~Nguyen, D.~Zan, Z.~Lin, J.-G. Lou, and W.~Chen, ``Codet: Code generation with generated tests,'' \emph{arXiv preprint arXiv:2207.10397}, 2022.

\bibitem{islam2024mapcoder}
M.~A. Islam, M.~E. Ali, and M.~R. Parvez, ``Mapcoder: Multi-agent code generation for competitive problem solving,'' \emph{arXiv preprint arXiv:2405.11403}, 2024.

\bibitem{wei2024magicoder}
Y.~Wei, Z.~Wang, J.~Liu, Y.~Ding, and L.~Zhang, ``Magicoder: Empowering code generation with oss-instruct,'' in \emph{Forty-first International Conference on Machine Learning}, 2024.

\bibitem{silva2023repairllama}
A.~Silva, S.~Fang, and M.~Monperrus, ``Repairllama: Efficient representations and fine-tuned adapters for program repair,'' \emph{arXiv preprint arXiv:2312.15698}, 2023.

\bibitem{luo2023wizardcoder}
Z.~Luo, C.~Xu, P.~Zhao, Q.~Sun, X.~Geng, W.~Hu, C.~Tao, J.~Ma, Q.~Lin, and D.~Jiang, ``Wizardcoder: Empowering code large language models with evol-instruct,'' \emph{arXiv preprint arXiv:2306.08568}, 2023.

\bibitem{ji2023ai}
J.~Ji, T.~Qiu, B.~Chen, B.~Zhang, H.~Lou, K.~Wang, Y.~Duan, Z.~He, J.~Zhou, Z.~Zhang \emph{et~al.}, ``Ai alignment: A comprehensive survey,'' \emph{arXiv preprint arXiv:2310.19852}, 2023.

\bibitem{tang2024understanding}
Y.~Tang, D.~Z. Guo, Z.~Zheng, D.~Calandriello, Y.~Cao, E.~Tarassov, R.~Munos, B.~{\'A}. Pires, M.~Valko, Y.~Cheng \emph{et~al.}, ``Understanding the performance gap between online and offline alignment algorithms,'' \emph{arXiv preprint arXiv:2405.08448}, 2024.

\bibitem{schulman2017proximal}
J.~Schulman, F.~Wolski, P.~Dhariwal, A.~Radford, and O.~Klimov, ``Proximal policy optimization algorithms,'' \emph{arXiv preprint arXiv:1707.06347}, 2017.

\bibitem{rafailov2024direct}
R.~Rafailov, A.~Sharma, E.~Mitchell, C.~D. Manning, S.~Ermon, and C.~Finn, ``Direct preference optimization: Your language model is secretly a reward model,'' \emph{Advances in Neural Information Processing Systems}, vol.~36, 2024.

\bibitem{ethayarajh2024kto}
K.~Ethayarajh, W.~Xu, N.~Muennighoff, D.~Jurafsky, and D.~Kiela, ``Kto: Model alignment as prospect theoretic optimization,'' \emph{arXiv preprint arXiv:2402.01306}, 2024.

\bibitem{steenhoek2023reinforcement}
B.~Steenhoek, M.~Tufano, N.~Sundaresan, and A.~Svyatkovskiy, ``Reinforcement learning from automatic feedback for high-quality unit test generation,'' \emph{arXiv preprint arXiv:2310.02368}, 2023.

\bibitem{dou2024stepcoder}
S.~Dou, Y.~Liu, H.~Jia, L.~Xiong, E.~Zhou, J.~Shan, C.~Huang, W.~Shen, X.~Fan, Z.~Xi \emph{et~al.}, ``Stepcoder: Improve code generation with reinforcement learning from compiler feedback,'' \emph{arXiv preprint arXiv:2402.01391}, 2024.

\bibitem{shen2023pangu}
B.~Shen, J.~Zhang, T.~Chen, D.~Zan, B.~Geng, A.~Fu, M.~Zeng, A.~Yu, J.~Ji, J.~Zhao \emph{et~al.}, ``Pangu-coder2: Boosting large language models for code with ranking feedback,'' \emph{arXiv preprint arXiv:2307.14936}, 2023.

\bibitem{shen2023large}
T.~Shen, R.~Jin, Y.~Huang, C.~Liu, W.~Dong, Z.~Guo, X.~Wu, Y.~Liu, and D.~Xiong, ``Large language model alignment: A survey,'' \emph{arXiv preprint arXiv:2309.15025}, 2023.

\bibitem{jelinek1977perplexity}
F.~Jelinek, R.~L. Mercer, L.~R. Bahl, and J.~K. Baker, ``Perplexity—a measure of the difficulty of speech recognition tasks,'' \emph{The Journal of the Acoustical Society of America}, vol.~62, no.~S1, pp. S63--S63, 1977.

\bibitem{miaschi2021makes}
A.~Miaschi, D.~Brunato, F.~Dell’Orletta, and G.~Venturi, ``What makes my model perplexed? a linguistic investigation on neural language models perplexity,'' in \emph{Proceedings of Deep Learning Inside Out (DeeLIO): The 2nd Workshop on Knowledge Extraction and Integration for Deep Learning Architectures}, 2021, pp. 40--47.

\bibitem{hu2021lora}
E.~J. Hu, Y.~Shen, P.~Wallis, Z.~Allen-Zhu, Y.~Li, S.~Wang, L.~Wang, and W.~Chen, ``Lora: Low-rank adaptation of large language models,'' \emph{arXiv preprint arXiv:2106.09685}, 2021.

\bibitem{rong2024distilling}
G.~Rong, Y.~Yu, Y.~Zhang, H.~Zhang, H.~Shen, D.~Shao, H.~Kuang, M.~Wang, Z.~Wei, Y.~Xu \emph{et~al.}, ``Distilling quality enhancing comments from code reviews to underpin reviewer recommendation,'' \emph{IEEE Transactions on Software Engineering}, 2024.

\bibitem{pearson1900x}
K.~Pearson, ``X. on the criterion that a given system of deviations from the probable in the case of a correlated system of variables is such that it can be reasonably supposed to have arisen from random sampling,'' \emph{The London, Edinburgh, and Dublin Philosophical Magazine and Journal of Science}, vol.~50, no. 302, pp. 157--175, 1900.

\bibitem{brown2020language}
T.~Brown, B.~Mann, N.~Ryder, M.~Subbiah, J.~D. Kaplan, P.~Dhariwal, A.~Neelakantan, P.~Shyam, G.~Sastry, A.~Askell \emph{et~al.}, ``Language models are few-shot learners,'' \emph{Advances in neural information processing systems}, vol.~33, pp. 1877--1901, 2020.

\bibitem{papineni2002bleu}
K.~Papineni, S.~Roukos, T.~Ward, and W.-J. Zhu, ``Bleu: a method for automatic evaluation of machine translation,'' in \emph{Proceedings of the 40th annual meeting of the Association for Computational Linguistics}, 2002, pp. 311--318.

\bibitem{shypula2023learning}
A.~Shypula, A.~Madaan, Y.~Zeng, U.~Alon, J.~Gardner, M.~Hashemi, G.~Neubig, P.~Ranganathan, O.~Bastani, and A.~Yazdanbakhsh, ``Learning performance-improving code edits,'' \emph{arXiv preprint arXiv:2302.07867}, 2023.

\bibitem{tihanyi2023formai}
N.~Tihanyi, T.~Bisztray, R.~Jain, M.~A. Ferrag, L.~C. Cordeiro, and V.~Mavroeidis, ``The formai dataset: Generative ai in software security through the lens of formal verification,'' in \emph{Proceedings of the 19th International Conference on Predictive Models and Data Analytics in Software Engineering}, 2023, pp. 33--43.

\bibitem{croft2023data}
R.~Croft, M.~A. Babar, and M.~M. Kholoosi, ``Data quality for software vulnerability datasets,'' in \emph{2023 IEEE/ACM 45th International Conference on Software Engineering (ICSE)}.\hskip 1em plus 0.5em minus 0.4em\relax IEEE, 2023, pp. 121--133.

\bibitem{liu2019roberta}
Y.~Liu, M.~Ott, N.~Goyal, J.~Du, M.~Joshi, D.~Chen, O.~Levy, M.~Lewis, L.~Zettlemoyer, and V.~Stoyanov, ``Roberta: A robustly optimized bert pretraining approach,'' \emph{arXiv preprint arXiv:1907.11692}, 2019.

\bibitem{biderman2024lora}
D.~Biderman, J.~G. Ortiz, J.~Portes, M.~Paul, P.~Greengard, C.~Jennings, D.~King, S.~Havens, V.~Chiley, J.~Frankle \emph{et~al.}, ``Lora learns less and forgets less,'' \emph{arXiv preprint arXiv:2405.09673}, 2024.

\end{thebibliography}

\end{document}